\documentclass[twocolumn,showpacs,preprintnumbers,amsmath,amssymb,prl]{revtex4}
\usepackage{graphicx}
\usepackage{dcolumn}

\usepackage[latin1]{inputenc}
\usepackage[T1]{fontenc}
\usepackage{ae,aecompl}

\newcommand{\beq}{\begin{equation}}
\newcommand{\eeq}{\end{equation}}
\newcommand{\bea}{\begin{eqnarray}}
\newcommand{\eea}{\end{eqnarray}}

\newcommand{\Rfg}[1]{Fig. \ref{F#1}}

\newcommand{\Rtb}[1]{Table \ref{T#1}}

\newcommand{\beginsupplement}{%
  \setcounter{table}{0}
  \renewcommand{\thetable}{S\arabic{table}}%
  \setcounter{figure}{0}
  \renewcommand{\thefigure}{S\arabic{figure}}%
}

\begin{document}

\author{Alexey K. Mazur}
\affiliation{
UPR9080 CNRS, Universit\'e Paris Diderot, Sorbonne Paris Cit\'e,\\
Institut de Biologie Physico-Chimique,\\
13, rue Pierre et Marie Curie, Paris,75005, France.}

\title[]{Homologous Pairing between Long DNA Double Helices}

\begin{abstract}
Molecular recognition between two double stranded (ds) DNA with homologous
sequences may not seem compatible with the B-DNA structure because the
sequence information is hidden when it is used for joining the two
strands. Nevertheless, it has to be invoked to account for various
biological data. Using quantum chemistry, molecular mechanics, and hints
from recent genetics experiments I show here that direct recognition
between homologous dsDNA is possible through formation of short
quadruplexes due to direct complementary hydrogen bonding of major groove
surfaces in parallel alignment. The constraints imposed by the predicted
structures of the recognition units determine the mechanism of
complexation between long dsDNA. This mechanism and concomitant
predictions agree with available experimental data and shed light upon the
sequence effects and the possible involvement of topoisomerase II in the
recognition.
\end{abstract}


\maketitle

Mutual recognition between dsDNA with identical sequences is a
long-standing enigma in molecular biology \cite{Kleckner:93}. It is
involved in processes including pre-meiotic and somatic paring of
homologous chromosomes \cite{McKee:04,Williams:07}, repeat-induced DNA
modifications \cite{Hagemann:96,Selker:02,Rossignol:94} and double strand
break repair \cite{Barzel:08}. Recognition is generally assumed to occur
similarly to homologous recombination, i.e., due to recruited proteins
that temporarily open dsDNA and make possible the cross-stranded
Watson-Crick (WC) base pairing. However, this would require proteins with
very special functions, whereas so far searches including genome-wide
genetic screens \cite{Bateman:08,Blumenstiel:08,Joyce:12} have not
revealed suitable candidates.  Direct DNA-DNA recognition has been
suggested as an alternative solution
\cite{Kleckner:93,McGavin:71,Wilson:79,Zickler:15}.  Two possible
mechanisms have been considered in the recent years: (1) Attractive long-range
electrostatic interactions between B-DNA with identical sequence-dependent
conformations \cite{Kornyshev:01,Kornyshev:10} and (2) a strand exchange
between two dsDNA to form the PX-DNA motif used in DNA-origami
nanotechnology \cite{Seeman:01,Wang:10}. These models explained available
biological data and fit well with the results of {\em in vitro}
experiments in cell-free conditions \cite{Baldwin:08,Danilowicz:09}.
However, they cannot account for the phenomenon of recognition between
partial homologies recently discovered by Gladyshev and Kleckner
\cite{Gladyshev:14}. These authors studied the sequence dependence of
repeat-induced point mutations (RIPs). RIPs occur in fungi cells that
somehow identify and target for mutation any long repeated sequence in
genome \cite{Hagemann:96,Selker:02}.  Strikingly, the
recognition occurs with even 25\% homology, provided that it is
distributed in a series of triplets spaced by 11 or 12 base pair steps
(bps) \cite{Gladyshev:14}. Two dsDNA with such sequences cannot form PX-DNA \cite{Shen:04b}
and neither they can be structurally similar, therefore, the RIP data
\cite{Gladyshev:14} do not fit with the mechanisms of direct recognition
\cite{Kornyshev:01,Kornyshev:10,Seeman:01,Wang:10}. These new observations
are also difficult to reconcile with any recognition via WC pairing.
Indeed, the RIP data indicate that the two dsDNA remain torsionally rigid
and the recognition improves with the number of active triplet frames
rather than the integral homology \cite{Gladyshev:14}. In contrast,
local melting should zero the twisting rigidity, and hybridization of
continuous homologous ssDNA should be orders of magnitude more efficient
than pairing of the same number of base pairs in periodically spaced
triplets.

In the present study I analyze the possibility of dsDNA recognition
through direct binding by major grooves. It has been noticed long ago
that the major groove edges of WC base pairs have complementary
hydrogen bonding (H-bonding) patterns \cite{Lowdin:64,Kubitschek:66}.
An infinite helical quadruplex using major groove association between
WC pairs was predicted by manual modeling \cite{McGavin:71} and
discussed as an intermediate state in homologous recombination
\cite{Wilson:79,Lebrun:95}. Experimentally, such structures were not
found, but the possibility of major groove H-bonding was confirmed
\cite{Kettani:95,Zhang:01}.  Using methods of quantum chemistry (QC),
molecular mechanics (MM), and molecular dynamics (MD), I show that
direct dsDNA binding by complementary major grooves should be
considered as a probable pathway for direct homology recognition.  The
admissible recognition conformations are dictated by structural
constraints and they explain experimental data better than alternative
mechanisms.

\begin{figure}[ht]
\centerline{\includegraphics[width=8.5cm]{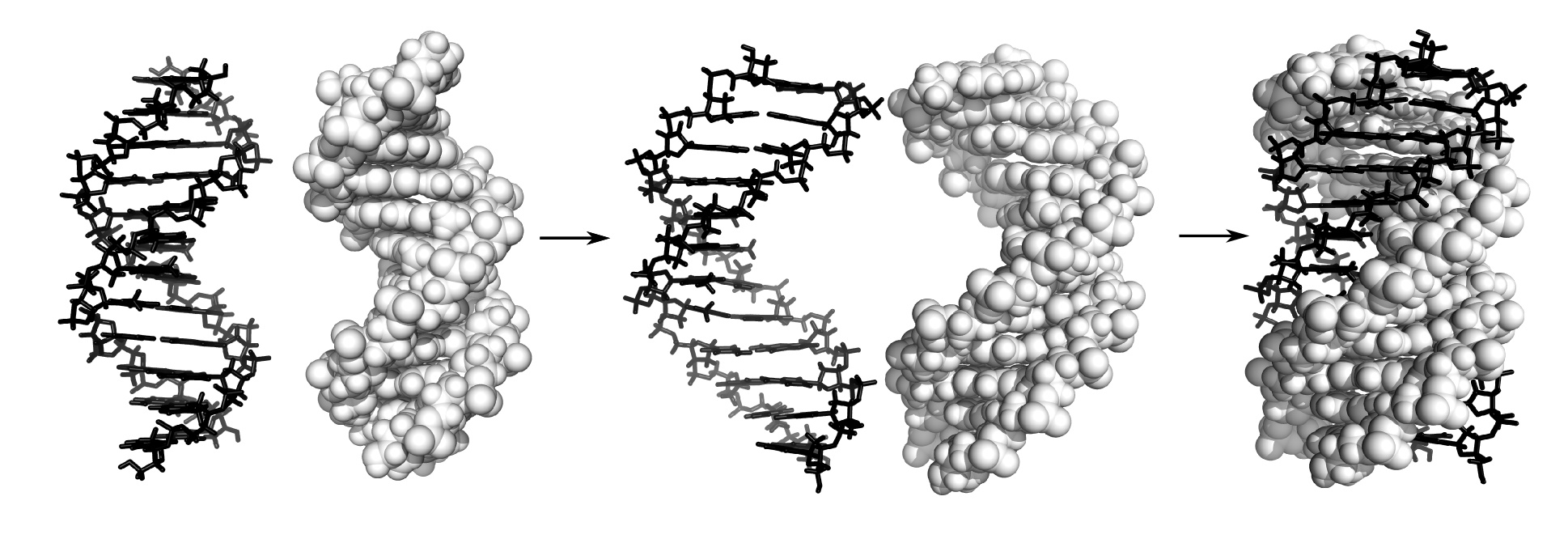}}
\caption{\label{Fmol1}
Helical quadruplex formed from two WC double helices. On the left, two
canonical B-DNA duplexes are shown facing one another by their major
grooves. The middle panel displays their conformations in the complexed
state. They are slightly stretched and untwisted to a helical pitch
about 12.9. The right panel shows a right-handed quadruplex formed by
major groove association.
}\end{figure}

\Rfg{mol1} illustrates the hypothetical association of two identical B-DNA
structures by merging the major grooves.  The quadruplex in the right
panel has four grooves, namely, two minor grooves of the constituting
double helices and two new grooves, hereafter called secondary, formed
between the two juxtaposed major grooves.  The interior of this structure
consists of parallel base tetrads shown in \Rfg{bp1}. They are formed by
identical WC pairs linked by two new H-bonds also called secondary. Such
tetrads were experimentally confirmed for both types of WC pairs using
short ssDNA hairpins \cite{Kettani:95,Zhang:01}. \Rfg{bp1} and \Rtb{bp1} reveal that
the secondary H-bonds are shorter and stronger than their sisters in
WC pairs. These results were obtained by QM optimizations of tetrad
geometries in vacuum \cite{Epaps}. Earlier studies of WC base pairs
indicate that such calculations reproduce experimental trends, with
the energy differences scaled down due to the polar environment
\cite{Sponer:04}. The stabilization energy of the GC/GC tetrad is
surprisingly large, namely, the energy of two H-bonds appears similar
to that of the WC pair with three H-bonds.  This non-pair-additive
electrostatic effect is due to the large dipole moment and high
polarizability of the GC pair \cite{Epaps}.  Because of this
non-pair-additivity, molecular mechanics significantly underestimates
the strength of secondary H-bonding, which is important for
interpretation of other results.

\begin{figure}[ht]
\centerline{\includegraphics[width=8.5cm]{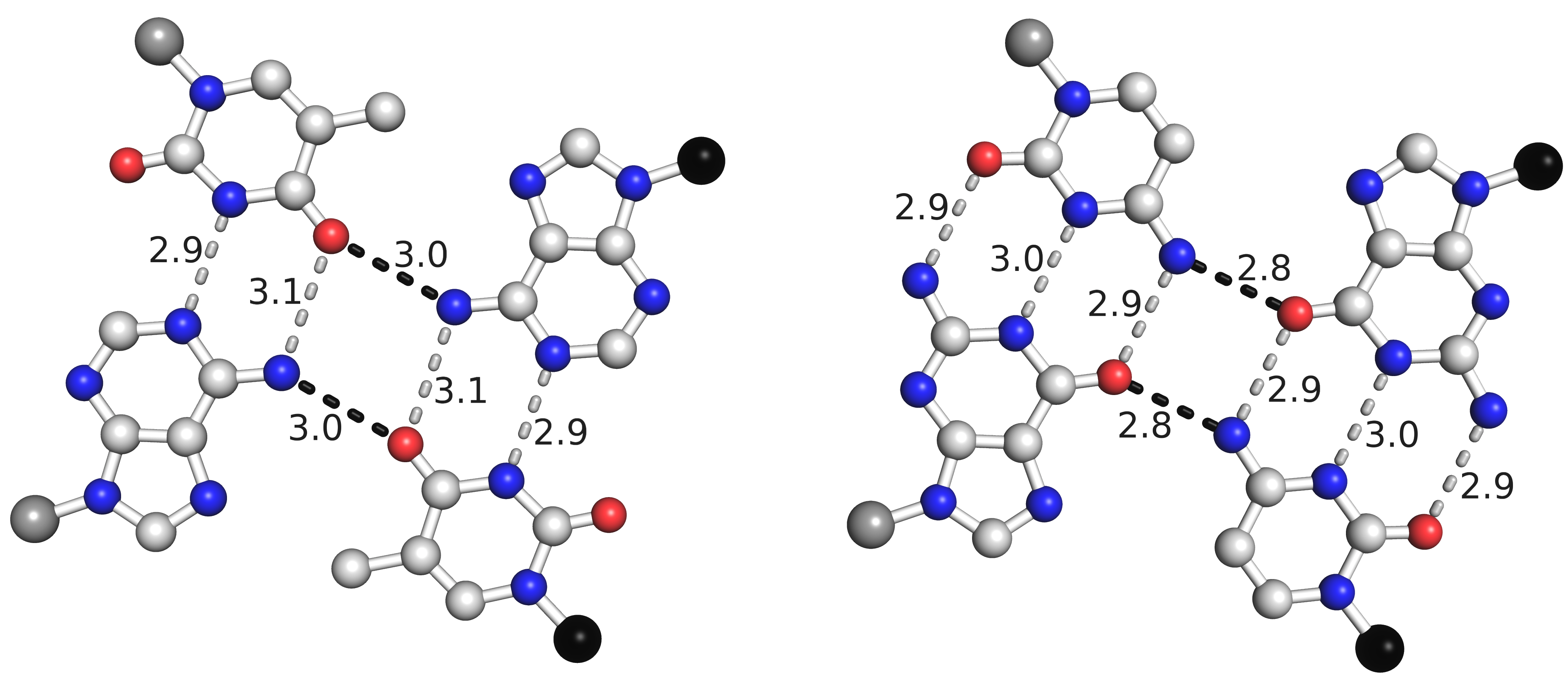}}
\caption{\label{Fbp1}(Color online)
Mutual recognition between identical WC base pairs via
major-groove edges. Large spheres of two different colors correspond to C1'
atoms of different dsDNA. Geometry and stabilization energies (\Rtb{bp1})
of individual base pairs and tetrads were evaluated by vacuum geometry
optimizations at MP2/6-311G(d) level of theory \cite{Epaps}. The computed
H-bond lengths are shown in angstroms (\AA). The secondary H-bonds are
distinguished by darker dashed lines.
}\end{figure}

\begin{table}[ht]\caption{\label{Tbp1}
Vacuum stabilization energies, U (kcal/mol), computed by QC and MM
methods as described elsewhere \cite{Epaps}. The energy of WC pairing
was estimated as the difference between the vacuum energy of the pair
and that of constituting nucleobases. For secondary pairing the energy
is obtained as the difference between the tetrad energy and that of
constituting base pairs.  The WC pairs are denoted by the standard
two-letter code. Slashes denote the secondary paring.  The $\rm
U_{MM}$ values were computed with the AMBER forcefield
\cite{Cornell:95}. For WC pairs these results agree with earlier data
\cite{Sponer:04}.}
\centerline{\begin{tabular}[t]{|c|c|c|c|c|}  \hline
Energy    & GC    & GC/GC & AT    & AT/AT \\ \hline                               
$\rm U_{QC}$  & 29.5  &  26.4 & 16.2  & 12.6  \\
$\rm U_{MM}$  & 28.3  &  19.5 & 12.9  & 10.6  \\ \hline
\end{tabular}}
\end{table}

The B-DNA conformations in the left panel of \Rfg{mol1} look
predisposed for association because they resemble separated fibers
of a two-strand twine. In the complex, the double helices are spun so
that quadruplexes longer than one turn cannot fall apart even in the
absence of the secondary H-bonding.  The complex is easily built by
making a cylinder from stacked tetrads and then properly placing
backbone strands at its surface \cite{McGavin:71}, but it cannot be
obtained by docking two dsDNA following the arrows in \Rfg{mol1}. To
this end, the two initial structures must be untwisted to an almost
flat ladder, joined, and then relaxed.  Even small
untwisting of dsDNA leads to dissociation of the two strands
\cite{Bryant:03}, therefore, this simple pathway is not feasible.  The
question is if there exists an alternative pathway that can join the
left and right hand states in \Rfg{mol1}. To get an idea of the
transition state of such a pathway, preliminary all-atom MD simulations were
run with quadruplexes of different lengths and sequences with explicit
ions and water \cite{Epaps}. In the course of these tests, it became
clear that a required transition state can be obtained by separating 5 bps
at both ends of the quadruplex, which gives four B-DNA "paws" protruding from
the core, and then keeping the paws wide open for the time necessary to
relax the helical twist to that of B-DNA.

The idea of the following MD simulations is similar to some early
MD studies \cite{Harvey:93} and it is explained in \Rfg{tj2}.
In all subsequents modeling only GC-alternating sequences were used. We start
from a predicted barrier state and try to reach both quadruplex and
unbound states in free dynamics, without any guiding restraints.  If we
are lucky the trajectories in both directions would go downhill on the
energy landscape. Even in this case, however, a straightforward simulation
requires enormous time resources.  Therefore, to obviate entropic
barriers, a Maxwell demon approach is applied. A bundle of trajectories is
started from the same state with different random velocities. All
trajectories are followed visually and stopped after a certain time
interval (usually about 2 ns) or when an interesting local transition
towards dissociation/binding occurred somewhere within the bundle. One of
the final states considered as most advanced towards dissociation/folding
is selected and used as the start of a new bundle of trajectories.  After
several iterations in the two opposite directions one gets two
trajectories leading to unbound and quadruplex states, respectively. The
trajectories are continuous in the coordinate space, with velocities
periodically randomized. By inverting one of them we obtain a pathway
between the terminal states in \Rfg{mol1} that involves only elastic
deformations and does not require base pair opening.

\begin{figure}[ht]
\centerline{\includegraphics[width=8.5cm]{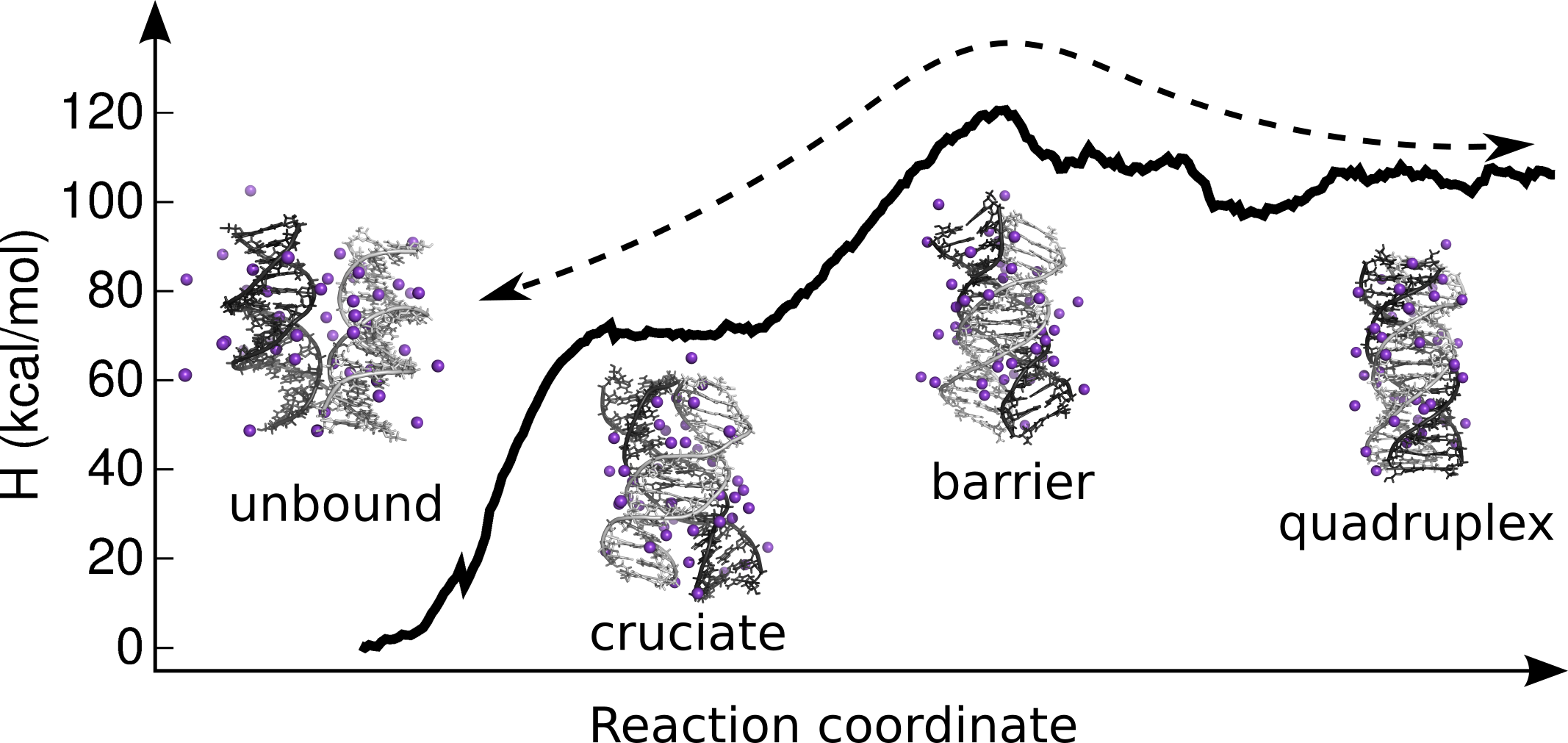}}
\caption{\label{Ftj2}
The overall plan of MD simulations and the approximate energy profile
along the transition pathway. Trajectories started from a predicted
barrier structure and continued in the two opposite directions. The
reaction coordinate was constructed as explained in the text, with the
energy profile smoothed by averaging with a sliding window of about 2
ns.
}\end{figure}

\begin{figure}[ht]
\centerline{\includegraphics[width=7.5cm]{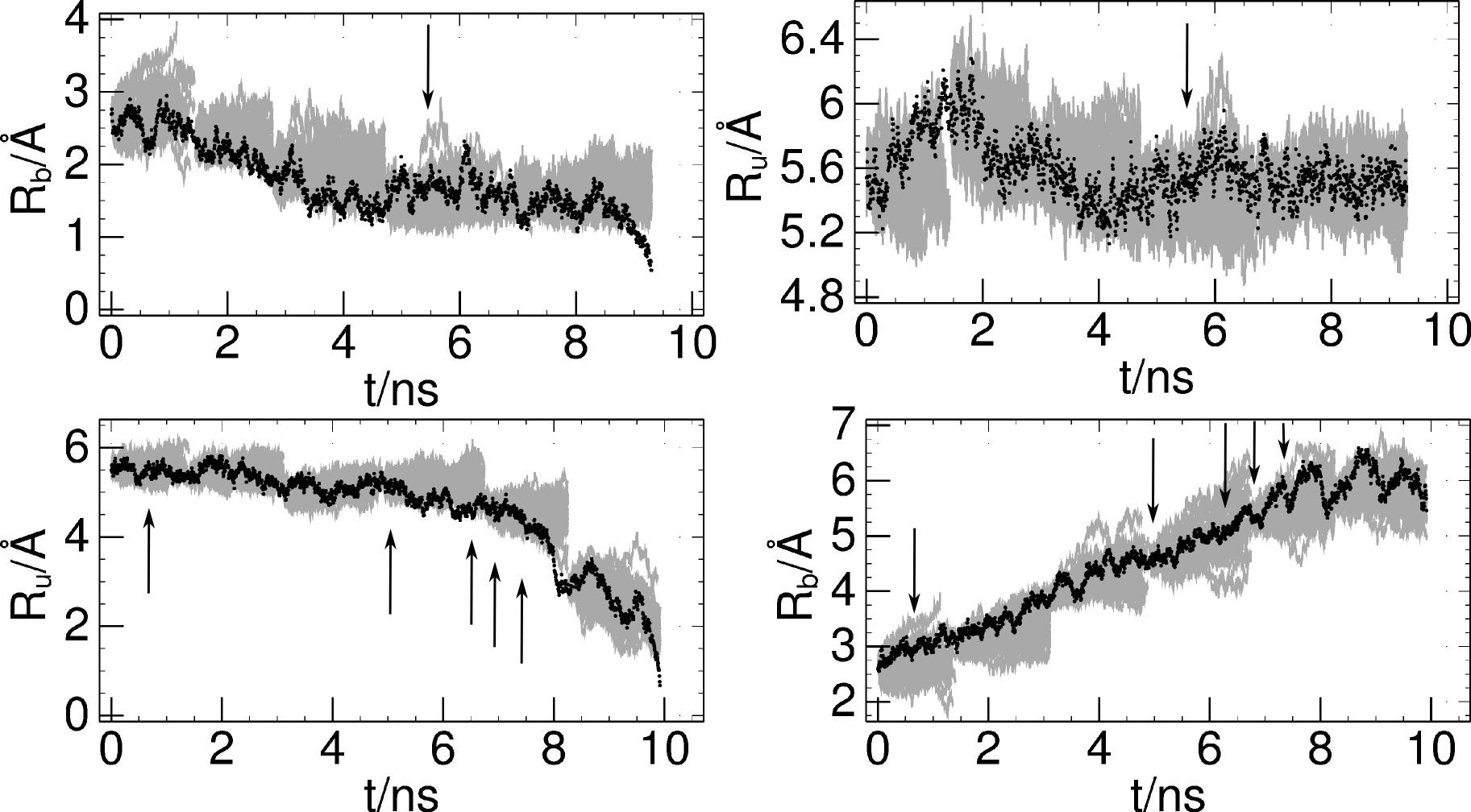}}
\caption{\label{Ftj1}
Time traces of root mean square deviations (RMSD). The RMSDs from
bound (quadruplex) (R$_b$) and unbound (R$_u$) states (see \Rfg{tj2})
were computed for two double helices separately and averaged. The
upper and lower panels display results for binding and dissociation,
respectively. The gray bands are formed by traces of bundles of 32
trajectories computed as explained in the text. The restart points can
be distinguished by narrowings of these bands. The black traces
correspond to trajectories selected for continuation. The vertical
arrows mark formation and splitting of tetrads during folding and
dissociation, respectively.
}\end{figure}

The time course of the production run is illustrated in \Rfg{tj1}. A
more detailed picture is provided as animation files \cite{Epaps}.
Surprisingly, just 10 ns were necessary to reach both quadruplex and
unbound states. During binding, only one complete additional tetrad
was formed. A few ions and water molecules were sequestered between
bases, which strongly complicated formation of secondary H-bonds.
However, both secondary grooves, with characteristic chains of
potassium ions between phosphates, were already formed. During
dissociation the order of events was inverted, that is, the
dissociation of tetrads preceded that of ions and groove opening. The
starting intermediate was probably shifted towards the quadruplex
state.  None of the trajectories of the first bundle displayed strong
trends towards dissociation even though in two cases one boundary
tetrad was split. The first of these trajectories selected for
continuation towards the unbound state was not successful. The second
choice worked; however, even in the fourth bundle there were
trajectories that turned back to folding (see \Rfg{tj1}). The
dissociation accelerated after the split of three tetrads.

The energy profile shown in \Rfg{tj2} was evaluated as follows. New
trajectories were re-started similarly to the main run from 280 states
equally spaced in time along the transition pathway, and the average
total energy was evaluated for 0.5 ns after short equilibration. This
profile is approximate and lacks the entropic contribution of the free
energy, but it gives an estimate of shape and the order of magnitude
of the values involved. Thorough calculations using umbrella sampling
and the weighted histogram analysis would be prohibitively costly, and
they usually give qualitatively similar profiles scaled down by
one-two orders of magnitude \cite{Bock:13}.
The apparently large energies
obtained are not prohibitive. First, all available data indicate that
the recognition requires long incubation stages from hours to weeks,
therefore, the corresponding free energy barrier can well reach 10-20
kcal/mol. Second, the plateau at 70 kcal/mole mainly depends upon the
type and concentration of ions. The neutralizing amount of monovalent
ions used here was not meant to reproduce real conditions that 
probably involve a combination of mono and divalent ions with higher
concentrations.  Finally, the energy values in \Rfg{tj2} would be much
larger for alternative recognition models that include strand
dissociation.

The plateau at 70 kcal/mole in \Rfg{tj2} indicates that structures
with 3-4 stacked tetrads can represent a metastable state with a local
free energy minimum. In short quadruplexes the tetrads are propeller
twisted and slightly non-parallel, which allows the paws protruding
from the core to be separated without strong bends. With the length
of the tetrad stack
increased, the tetrads become more parallel and stronger
bending in the paws is required.  This explains the emergence of the
plateau in \Rfg{tj2} that can well transform into an energy minimum
with stronger secondary H-bonds corresponding to \Rtb{bp1}. I suggest
that, under appropriate ionic conditions, the free energy in this
minimum is lower than that of the unbound state. In contrast, the
additional bending strain responsible for the central energy barrier
cannot be eliminated. With increased DNA length this barrier would
broaden and eventually become a plateau.  Under these assumptions,
the cruciate structures with short quadruplexes of 3-4 stacked tetrads
work as recognition units in homologous alignment of long double
helices.

\begin{figure}[ht]
\centerline{\includegraphics[width=8.5cm]{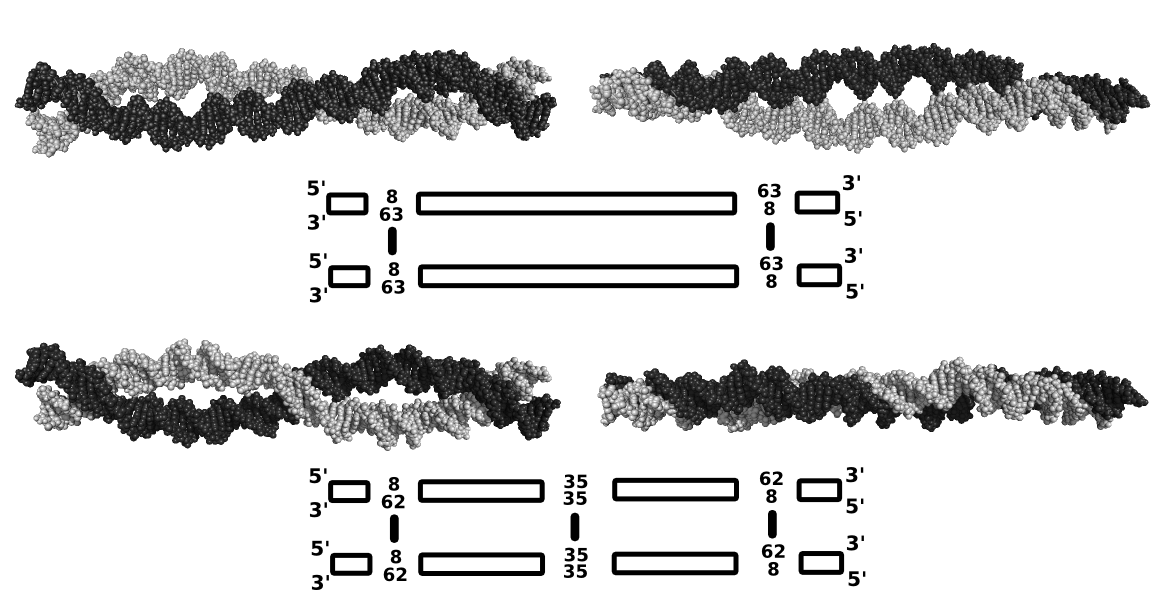}}
\caption{\label{Fmol2}
Modeled complexes of long dsDNA obtained as explained in the text.
The upper and lower panels demonstrate paranemic and plectonemic modes
of binding, respectively. The two columns show two perpendicular views of
each structure. In the schematics below the structures, dsDNA stretches are
shown as rectangles interrupted by recognition units.  For each unit the
four bases of the central tetrad are indicated, with the secondary
H-bonding marked by thick vertical lines.
}\end{figure}

Conformations of the cruciate units observed in dynamics were used for
predicting the complexes of long DNA (\Rfg{mol2}). They were built in
several steps by combining all-atom and coarse grained modeling
\cite{Epaps}. The cruciate shape of the recognition unit and the
bending rigidity of the double helix impose significant limitations
upon the minimal axial separation between the pairing contacts. In the
upper panels of \Rfg{mol2} the two recognition units are separated by
five helical turns. This number is identical in the two chains,
therefore, the units share the same plane. For smooth connection
the intermediate helices take a particular sinuous shape.
The high bending rigidity of DNA straightens these helices and pushes
them close to each other against repulsive electrostatic forces. The
helix-helix orientations are not optimal according to earlier
predictions and simulations \cite{Varnai:10,Cortini:11} and the
corresponding energy contributes to the high energy barrier of binding.
Even though the bend angles in the helical stretches are admissible, they
are strained because all local bends must have concerted
values and directions. With the growing distance between the
recognition units, this strain is relieved and its energy can be
compensated by that of binding. For complete homologies, the
experimental recognition threshold of 200-300 bp probably corresponds
to a relatively small number of specific contacts, which explains the
drastic disappearance of recognition at these DNA lengths.

The upper and lower structures in \Rfg{mol2}, respectively,
demonstrate paranemic and plectonemic contacts between two dsDNA. In
the latter case the linear density of recognition units is two times
higher. Plectonemes like that in \Rfg{mol2} are possible when the two
terminal recognition units are separated by an odd number of helical
turns. In this case the middle unit is rotated by 180$^\circ$ and
shifted by a half-integer number of turns. In a similar plectoneme
with an even number of helical turns the central segments of the two
double helices would face one another by their minor grooves. In this
case, one can consider the possibility of a cruciate recognition unit
formed by the hypothetical strand exchange mechanism \cite{Wilson:79}.
The paranemic contacts can be always formed by loops protruding from
chromosomes. In contrast, plectonemic contacts require the
topoisomerase II (Top2) activity because otherwise every right-hand
turn with three recognition units must be compensated by a left-hand
turn, with no recognition units and a very high entropic penalty.
Interestingly, the loss or inhibition of Top2 was shown to partially
compromise the pairing of homologous chromosomes in cell cultures
\cite{Williams:07}. It would be interesting to check if this effect
also plays a role in RIP.

Recognition via discrete units spaced by several helical turns
represents an extension of one of the models considered by Gladyshev
and Kleckner \cite{Gladyshev:14} and it sheds new light upon their
data. Notably, the helical pitch of the quadruplex structure in
\Rfg{mol1} (12.9) is larger than that of B-DNA (10.5). Therefore, the
average helical pitch in complexes shown in \Rfg{mol2} grows with the
density of recognition units, which explains the higher RIP efficiency
for sequences with periods 11 and 12 bp compared to 10 bp. The
large distance between the recognition units makes the constraint upon
the concerted twisting in the two double helices more stringent
because the amplitude of thermal torsional fluctuations grows only as
a square root of the chain length. Under normal temperature,
the difference between 10 and 11-bp periodicities can be
compensated by thermal fluctuations for one helical turn, but not for
five helical turns. In the latter case it corresponds to rotation by
180$^\circ$. This explains strong differences in RIP activities for
the some periodicities that differ by only one bp. Finally, the
sequence dependence of the secondary H-bonding predicted by \Rtb{bp1}
might account for the examples of strongly different RIP for
homologies that differ only by the sequence \cite{Gladyshev:14}.

Encounters between identical DNA sequences are rare in nature, but
they should be very frequent {\em in vitro}, therefore, one may ask
why complexes shown in \Rfg{mol2} remain almost unnoticed in chemical
laboratories. In fact, they are perhaps long known, but discarded. DNA
is never stored during hours and weeks in ionic solutions because it
is known to slowly aggregate and deteriorate. Rare attempts to
systematically study the slow evolution of DNA samples in laboratory
conditions gave very perplexing results \cite{Brewood:10}. The
sequence-specific association is driven by ions, with both mono and
divalent cations probably involved.  The binding shown in \Rfg{mol2}
occurs due to reversible interactions that are likely to be destroyed during
dilution or penetration through gels. At the same time, small ion
excess may lead to almost irreversible non-specific complexation.
These issues are not easy to sort out and they require further
experimental investigation.

In summary, the mutual recognition between two homologous B-DNA might
occur due to direct complementary H-bonding of major groove surfaces in
parallel alignment. The pairing of two dsDNA results in formation of a
planar cross-shaped recognition unit, with a central quadruplex of 3-4 bps
and four B-DNA paws protruding in opposite directions. In a complex of two
dsDNA the recognition units have to be spaced by at least several helical
turns, therefore, the binding requires long double helices, but only
partial homology. The recognition units are separated from the unbound
state by a high energy barrier and they are stabilized by specific
H-bonding as well as ion-DNA interactions. Therefore, the binding takes
very long time and is very sensitive to ionic conditions. The proposed
mechanism and concomitant predictions agree with earlier data
and shed light upon the recent intriguing experimental results \cite{Gladyshev:14}.

\begin{acknowledgments}
The computational resources used in this study were supported by the
Initiative d'Excellence program from the French State Grant No.
ANR-11-LABX-0011-01 (DYNAMO).
\end{acknowledgments}

\bibliography{last}

\clearpage

\section{Supplemental Material}
\setcounter{figure}{0}

\beginsupplement

\subsection*{Quantum chemistry and molecular mechanics calculations}

The chemical geometry and stabilization energies of WC pairs and
tetrads corresponding to major groove binding of two B-DNA were
evaluated by vacuum geometry optimizations at MP2/6-311G(d) level of
theory. This accuracy is similar to that in earlier studies
for nucleobases and base pairs \cite{Sponer:04,Srinivasan:09}.  The
Firefly QC package \cite{Granovsky:15} was used, which is partially
based on the GAMESS (US) \cite{Schmidt:93} source code.  For
comparison, similar energy minimization were carried out using
standard MM approximations with the all-atom AMBER forcefield
\cite{Cornell:95}.  AMBER is known to reasonably reproduce the quantum
energies of WC base pairing in vacuum \cite{Sponer:04}. To my
knowledge, for the secondary H-bonds this was never checked.

The comparison makes sense and is necessary only for nearly planar
configurations because, in dsDNA complexes of interest, non-planar
configurations are sterically impossible. In vacuum, however,
unconstrained energy minimizations starting from planar configurations
for tetrads in Fig.1 usually result in buckling because the out-of-plane
deformations are virtually free, whereas non-planar configurations
have lower energies due to additional atom-atom contacts. Moreover,
MP2-minimizations of guanine and cytosine give non-planar pyramidal
conformations of the exocyclic amino groups, which is not reproduced
in MM. For consistent comparison, therefore, the energy minimizations
were carried out with the overall planarity imposed by constraints. To
this end, the structures of bases were described using natural
internal coordinates, that is, bond lengths, bond angles and
dihedrals. Bases in WC pairs and tetrads were linked by virtual bonds.
The relative orientation of two linked bases is described by six
internal coordinates corresponding to virtual bonds and angles formed
at the junction. The set of six junction coordinates involves one
distance, two planar angles, and three dihedrals. The overall
planarity is imposed by freezing the dihedrals. The junctions were
chosen so that the free planar angles could not come close to zero or
$\pi$ during minimizations. In total, nine junction dihedrals were
frozen to impose the planarity of tetrads. In these conditions, the
exocyclic amino groups appeared to remain planar without additional
constrains. The constrained minimizations were carried out using the
built-in options of the Firefly QC package \cite{Granovsky:15}.
The results are shown in \Rtb{bp1}.

\begin{table}[h]\caption{\label{TSbp1}
Computed stabilization energies, U (kcal/mol), and dipole moments, D
(debye), of planar nucleobase complexes. The energy of WC pairing was
estimated as the difference between the vacuum energy of the pair and
that of constituting isolated nucleobases, respectively. For secondary
pairing the energy is obtained as the difference between the tetrad
energy and that of constituting base pairs, respectively. The WC pairs
are denoted by the standard two-letter code. Slashes denote the
secondary paring.}
\vskip 0.2cm
\centerline{\begin{tabular}[t]{|c|c|c|c|c|} \hline
 Complex
          & $U_{\rm QC}$    &   $D$    & $U_{\rm MM}$  \\
\hline                                             
    A     &   -         &  2.55    &   -       \\
    T     &   -         &  4.03    &   -       \\
    AT    &  16.2       &  1.27    &  12.9     \\
  AT/AT   &  12.6       &  0.00    &  10.6     \\
    G     &   -         &  6.61    &   -       \\
    C     &   -         &  6.27    &   -       \\
    GC    &  29.5       &  6.09    &  28.3     \\
  GC/GC   &  26.4       &  0.00    &  19.5     \\
\hline
\end{tabular}}
\end{table}

For WC pairs, both the QC and MM energies in \Rtb{Sbp1} are close to
those reported earlier \cite{Sponer:04}, which means that these values
are not very sensitive to the planarity constraints and other
methodological differences from earlier studies.  The dipole moments
also included in \Rtb{Sbp1} are important because dipole-dipole
interactions often give significant electrostatic contributions to the
stabilization energies. Notably, the dipole moment of the AT pair is
significantly smaller than those of adenine and thymine bases,
indicating that the corresponding vectors in the pair are
antiparallel, which increases the energy of pairing. This interaction
can involve a significant non-pair-additive component because
antiparallel molecular dipoles induce additional electron polarization
that enhances the electrostatic attraction. This explains the
significantly larger quantum stabilization energy of the AT pair
compared to MM (\Rtb{Sbp1}). In contrast, the large dipole moments of
guanine and thymine in the GC pair are strongly non-collinear, which
explains the much better agreement between the QC and MM stabilization
energies.

\begin{figure*}[ht]
\centerline{\includegraphics[width=15cm]{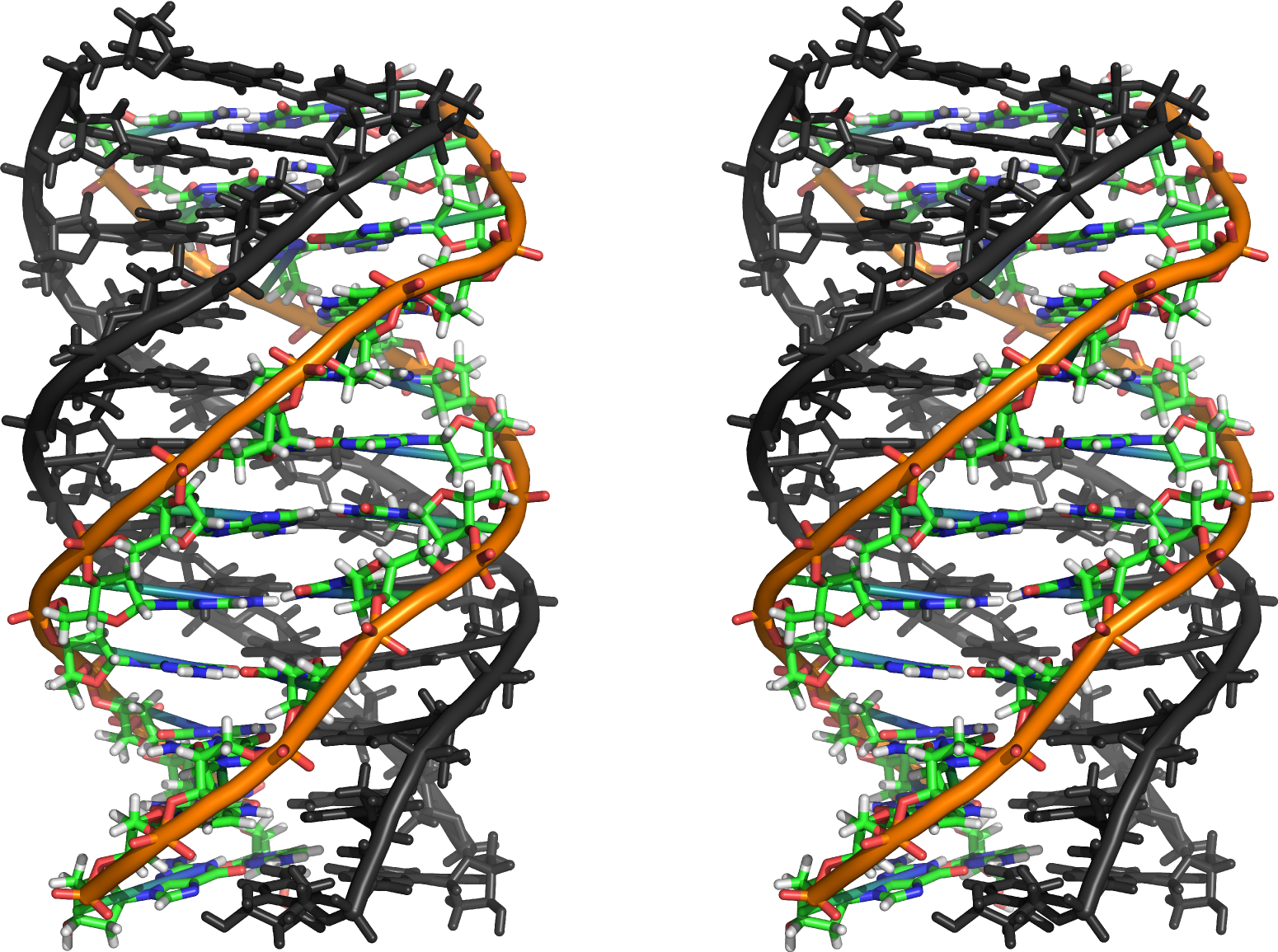}}
\caption{\label{Fquad}
A snapshot form an MD trajectory of a quadruplex formed by two
dodecamer double helices with the self-complementary sequence
CGCGAATTCGCG. One of the helices is colored black for distinction. A
cross-eyed stereo image is shown.
}\end{figure*}

Based upon the foregoing considerations one can anticipate a
significant contribution to the stabilization energy of tetrads from
the dipole-dipole polarization of constituting WC pairs. Indeed, the
tetrad dipole moment is zero by symmetry, that is, the dipole moments
of the constituting pairs are exactly antiparallel. The results shown
in \Rtb{Sbp1} confirm this prediction. For the AT/AT tetrad the QC and
MM stabilization energies are relatively close because the dipole
moment of the AT pair is small. In contrast, the quantum stabilization
energy of the GC/GC tetrad is strikingly large. Two secondary H-bonds
formed in the GC/GC tetrad give the stabilization energy similar to
that due to three H-bonds of the GC pair.

The stabilization energy of the AT/AT tetrad is somewhat smaller than
that of the AT pair, with a somewhat larger difference for $U_{\rm
QC}$ than for $U_{\rm MM}$. A part of this difference is due to two
steric clashes between the thymine methyls and N7 atoms of adenines in
the AT/AT tetrad (see Fig. 1). In the strictly planar configuration,
these contacts hinder formation of optimal geometry of secondary
H-bonds. When the energy minimization is continued without the
planarity constraints these clashes are readily relieved by small
propeller twisting, with the overall planarity perturbed
insignificantly. The corresponding energy gain is about 2 kcal/mole.
Further minimization leads to buckling of the tetrad and large-scale
rotation of base pairs toward a stacked configuration, which occurs
with very slow fall of energy.  Qualitatively similar results were
obtained for GC/GC tetrads with 5-methylcytosines (not shown). These
computational observations suggest that 5-methylation of pyrimidine
bases does not cause steric problems for homologous pairing between
dsDNA although further studies are required to take into account the
solvent and structural environment.

\section*{Molecular dynamics simulations}

DNA duplexes of different lengths and sequences were modeled in
aqueous environment neutralized by potassium ions \cite{Joung:08},
using a recent version of the all-atom AMBER forcefield
\cite{Cornell:95,Wang:00,Perez:07a,Zgarbova:13} with SPC/E water
\cite{Berendsen:87} in periodic boundaries. The electrostatic
interactions were treated by the SPME method \cite{Essmann:95}, with
the common values of Ewald parameters, that is 9 \AA\ truncation for
the real space sum and $\beta\approx 0.35$. The temperature was
maintained by the Berendsen algorithm \cite{Berendsen:84} applied
separately to solute and solvent with a relaxation time of 10 ps.  To
increase the time step, MD simulations were carried out by the
internal coordinate method (ICMD) \cite{Mzjcc:97,Mzjchp:99}, with the
internal DNA mobility limited to essential degrees of freedom. The
rotation of water molecules and internal DNA groups including only
hydrogen atoms was slowed down by weighting of the corresponding
inertia tensors \cite{Mzjacs:98,Mzjpc:98}. The double-helical DNA was
modeled with free backbone torsions as well as bond angles in sugar
rings, but rigid bases and phosphate groups.  The net effect of these
constraints upon DNA dynamics is not significant, which was checked
earlier through comparisons with conventional Cartesian MD
\cite{Mzjacs:98,Mzbj:06}. The time step was 0.01 ps.

Helical quadruplex conformations were constructed in several steps
using in-house software. Ideal quadruple helices were built from dsDNA
with homopolymer polyG-polyC sequences. To this end, two dsDNA trimers
were docked manually in vacuum and the structure was energy minimized
in internal coordinates with constraints ensuring that in each single
strand homologous internal coordinates at different steps had
identical values. The resulting structure was elongated by adding
tetrads one by one, with energy minimizations repeated from
conformations obtained in the previous step. The helical backbone
conformations obtained were used for building quadruplexes with other
sequences. The subsequent MD simulations were carried out using
standard equilibration and production protocols earlier applied to
dsDNA \cite{Mzprl:10,Mzpre:11}.

\begin{figure*}[ht]
\centerline{\includegraphics[width=16cm]{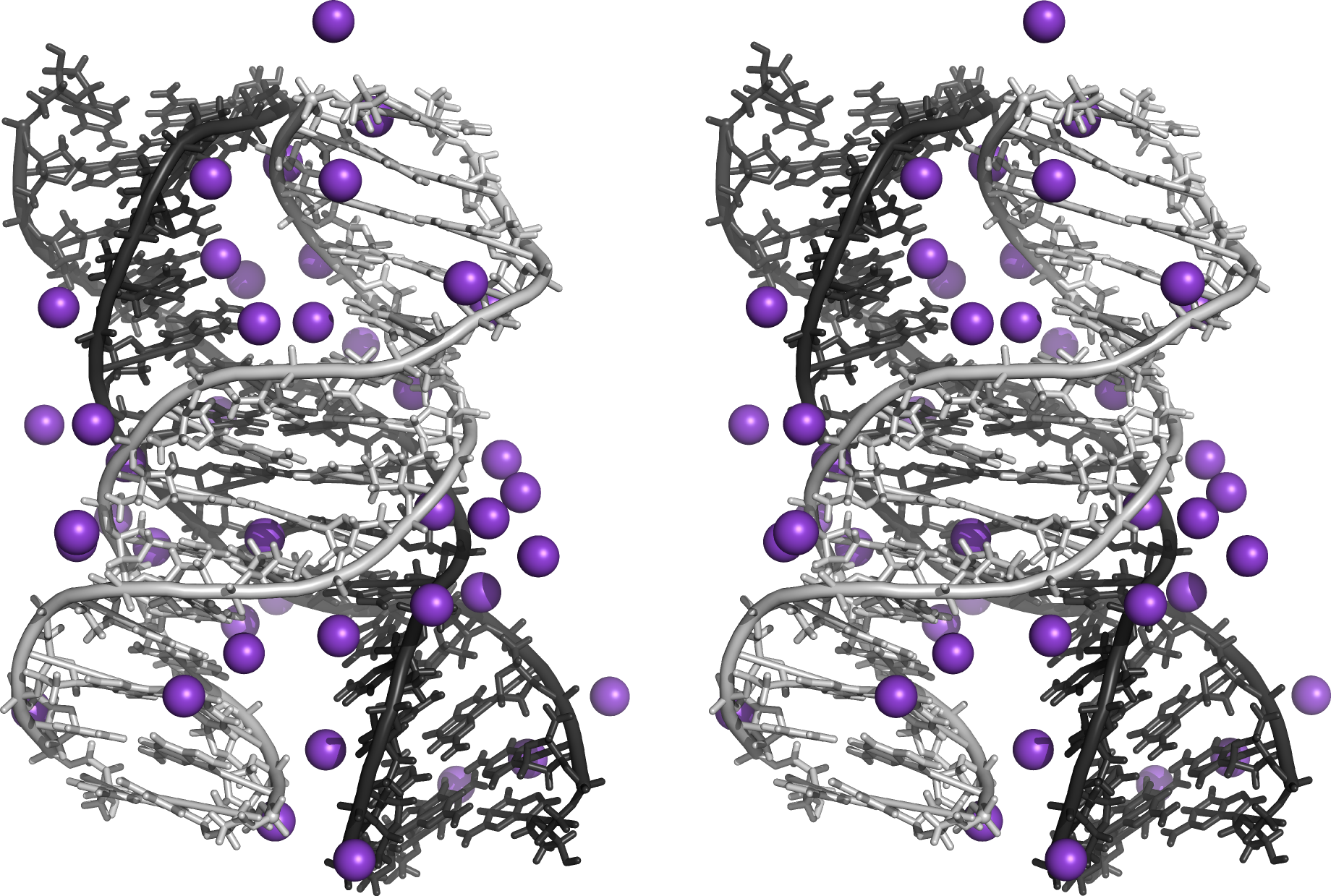}}
\caption{\label{Fcru}
A snapshot from the dissociation pathway with the cruciate structure
used for modeling complexes of long DNA presented in Fig. 5. The
solvent is not included except for potassium ions from the 10 \AA\
vicinity of the complex. A cross-eyed stereo image is shown.
}\end{figure*}

A few preliminary MD simulations involved helical quadruplexes formed
by double helical polyG-polyC decamers, polyGC pentadecamers and
dodecamers with the self-complementary sequence CGCGAATTCGCG. A
snapshot from the last trajectory is shown in \Rfg{quad}.  The
following qualitative computational observations seemed most
interesting for the present study. Originally \cite{McGavin:71}, the
quadruplex structure formed by merging two dsDNA seemed appealing
because of (i) the compactness and (ii) the complementarity of the
H-bonding. Both these features turned out to be inessential. In fact,
the secondary grooves remain hydrated and allow water and ions to
penetrate inside and even break the secondary H-bonds. This, however,
is not critical for stability because the secondary grooves are sealed
by a layer of ions sandwiched between the backbone phosphate groups.
In one test all the secondary HB in the CGCGAATTCGCG quadruplex were
canceled by adding appropriate repulsive forces, and this had
virtually no effect upon the stability in the nanosecond time range.
When the duplexes in the pentadecamer quadruplex are separated near
one end by properly applying perpendicular stretching forces the
secondary grooves are quickly closed once the stretching is switched
off, but the specific H-bonding usually is not recovered completely.

According to \Rtb{Sbp1} the AMBER forcefield significantly
underestimates the relative strength of the secondary H-bonding, which
certainly affected the above results. At the same time, the adsorption
of cations in the secondary grooves, evident in MD, is quite probable
and it can result in a strong electrostatic stabilization depending
upon ion types and concentrations. In the course of these tests, it
became clear that the transition state between the bound and
dissociated states of two dsDNA can be obtained by separating 5 bps at
both ends of the quadruplex, which gives four "paws" protruding from
the core, and then keeping the paws wide open for the time necessary
to relax the helical twist to that of B-DNA.  The preparation of such
intermediate turned out to be quite laborious because it could not be
achieved by vacuum modeling. Eventually, this was done by running MD
simulations of two pentadecamer dsDNA with GC-alternating sequences
under visual control with restraining potentials as well as
non-conservative forces adjusted {\em ad hoc} so that the structure
was slowly pushed towards the predicted state. The resulting structure
included a quadruplex core of five stacked tetrads in the middle and
four B-DNA paws of the same length protruding in different directions.
The WC pairing within the duplexes was intact.  The base stacking was
not strictly parallel, but there was no significant deformations
accompanied by water entering between bases.  The duplexes were
hydrated and made no direct contacts except the secondary H-bonds. All
backbone torsion angles had values within the canonical B-DNA
intervals. These computations involved many tens of nanoseconds of MD,
which also served for slow ion diffusion and equilibration of the
environment. Therefore, the subsequent production MD simulations
could be started after randomization of velocities followed by short
re-equilibration during a few tens of picoseconds.

\subsection*{Complexes of long DNA}

The structures of complexes of long DNA shown in Fig. 5 were built in
several steps by combining all-atom and coarse grained modeling. The
pentadecamer cruciate structures with three central tetrads and the
B-DNA paws that could be continued without strong bends were selected
from MD trajectories by visual inspection (see \Rfg{cru}). In the
subsequent modeling these structures remained rigid. Using the
standard local base pair frames \cite{Olson:01} the discrete wormlike
rod (WLR) versions \cite{Mzjpc:09} of these conformations were built.
The objective of the coarse-grained step was limited to construction
of chains that can be transformed into all-atom models without atom
clashes. To this end, an iterative trial-and-error procedure was used.
Rigid WLR models of cruciate units were fixed in space at certain
distances and joined by flexible WLR fragments \cite{Mzjpc:09} of
necessary lengths. The harmonic parameters of the WLR model were
earlier chosen so that in Monte Carlo and Brownian dynamics
simulations the chains were statistically close to B-DNA, namely, the
average twist 33.5$^\circ$, the average rise 3.3 \AA, the bending
persistence length 50 nm, and the torsional persistence length 90 nm.
These fragments were kept at distances of about 25 \AA\ by harmonic
restraints and the system was rapidly relaxed by short MD followed by
energy minimization. After that the helical parameters of the flexible
fragments were checked. The twist and rise values almost did not vary
along the flexible chains and the distances between the cruciate units
were changed by trials to push the corresponding values to the
harmonic minima. The final WLR models were back-transformed into all
atom structures. With the helical parameters close to B-DNA and the
absence of strong bends, the backbone junctions were easily corrected
by vacuum energy minimization, with position restraints applied to
bases only.

\subsection*{Trajectory animation files}
Files fold1.mp4 and fold2.mp4 show two perpendicular views of the
consecutive states from the trajectory starting from the predicted
intermediate state and continued to the bound state (folding). The
movies represent cross-eyed stereograms of DNA with potassium ions
from the 10 \AA\ vicinity of backbone atoms. Water molecules are not
shown.  Files unfold1.mp4 and unfold2.mp4 present similar data from
the trajectory starting from the same intermediate state and continued
to dissociation (unfolding). The animation files were prepared with
PyMOL program.


\end{document}